\documentclass[aps,preprint,epsfig,rotate]{revtex4}
\begin{document}
\title{On the interaction between two point electric charges}

 \author{Alexei M. Frolov}
 \email[E--mail address: ]{afrolov@uwo.ca}

\affiliation{Department of Applied Mathematics\\
 University of Western Ontario, London, Ontario N6H 5B7, Canada}



\date{\today}

\begin{abstract}

The general formula for the interaction potential between two point electric charges which 
contains the lowest order corrections to the vacuum polarization is derived and investigated. 
Analytical derivation of this formula is based on the closed analytical expression for the 
Uehling potential obtained earlier. Our analytical formula has the correct asymtotic behaviour 
at small and large distances between two interacting electric charges. We also discuss a number 
of problems which are of great interest in applications, e.g., vacuum polarization corrections 
to the Coulomb scattering and cusp value between two electrically charged particles. An original 
algebraic procedure developed in this study allows one to determine consequtive corrections for 
vacuum polarization for Coulomb systems.   

\end{abstract}

\maketitle
\newpage

\section{Introduction}

In classical electrodynamics the interaction potential $V$ between two point electric charges 
$q_1 e$ and $q_2 e$ is described by the Coulomb law, i.e. $V \simeq \frac{q_1 q_2 
e^2}{r_{12}}$, where $r_{12}$ is the interparticle distance. This expression for $V$ does not 
change its form in the non-relativistic Quantum Mechanics \cite{Heitl}, when the Planck constant 
$h$ is finite. However, if we also assume that the speed of light $c$ is finite, then Quantum 
Electrodynamics leads us to the necessaty to modify the Coulomb law. The main correction is 
related to the vacuum polarization in the spatial areas close to the electric charges. In the 
lowest order approximation such a correction is represented by the Uehling potential \cite{Uehl}, 
which is correct at short interparticle distances. However, the Uehling potential does not provide 
the correct long-range asymptotics. Briefly, the behaviour of the Uehling potential at large $r$ is 
not correct, since it decreases at large $r$ exponentially, i.e. too rapidly. On the other hand, it
can be shown that the electric field generated by the vacuum polarization has the long-range asymptotic 
which decreases at $r \rightarrow \infty$ much slower than exponential function. Therefore, the actual 
potential which describes the effects of vacuum polarization must be represented as a sum of the Uehling 
potential and an additional potential $W_K(r)$ which corrects the exponential decay of the Uehling 
potential at large $r$. Such a potential $W_K(r)$ was found by Wichmann and Kroll in \cite{WK}. 

Our goal in this study is to investigate the properties of the $U(r) + W_K(r)$ potential and evaluate 
the corresponding corrections produced by this potential in the solutions of the Schr\"{o}dinger
equation. The problem has substantial scientific and methodological interest. There are two ways which
can be used to evaluate these corrections. First, we can evaluate the expectaton value(s) of the $U(r) 
+ W_K(r)$ potential. In the second approach the additional $U(r) + W_K(r)$ term is directly introduced 
as an additional potentials in the Schr\"{o}dinger equation. By solving this Schr\"{o}dinger equation 
for the bound states we can investigate changes in many bound state properties produced by the vacuum
polarization corrections. The largest and most interesting changes can be found in the cusp values which
are determined for each pair of electrically charged particles. Vacuum polarization also contributes to 
the scattering of the two electrically charged particles.   

The goal of this study is to summarize all important details known for the Uehling and Wichmann-Kroll 
potentials which describe the lowest order correction for vacuum polarization in atomic systems. Some of these 
properties were considered in our earlier studies \cite{Our1}, \cite{Our2} for the Uehling potential. Here we 
want to re-derive and analyze the actual potential, i.e. the sum of the Uehling potential with some additional 
terms, which acts between two electrically charged quantum particles. As we mentioned above such a potential 
must be derived from Quantum Electrodynamics in the lowest order approximation upon the fine-structure constant 
$\alpha = \frac{e^2}{\hbar c} \approx \frac{1}{137}$. Moreover, it must have correct asymptotic behaviour at 
large and small interparticle distances. 

This study begins by a brief review (see, Sections II - IV) of the recent results derived for the potentials 
which describe the effects of vacuum polarization in the lowest order). Our analysis of vacuum polarization
performed in Sections V - VII includes a number of interesting results which were never discussed earlier. 
In particular, we consider here numerical changes in the Coulomb cusp values which are produced by the 
vacuum polarization. Another interesting problem is to detemine possible changes in the Coulomb scattering 
related with the vacuum polarization. We also develop an algebraic approach for analytical description of 
the vacuum polarization.  

\section{Uehling potential}

In our earlier study \cite{Our1} we have derived the closed analytical expression for the Uehling potential $U(r)$ 
which describes the lowest order correction (upon the fine structure constant $\alpha$) for vacuum polarization to 
the regular Coulomb potential. Formally, this correction is related to the polarization of the vacuum produced by a 
point electric charge. The total interaction potential $\phi(r)$ between two electric charges $e$ and $Q e$ is 
written in the form (in atomic units $\hbar = 1, m_e = 1$ and $e = 1$)
\begin{eqnarray}
 \phi(r) = \frac{Q}{r} + \frac{2 \alpha Q}{3 \pi r} \cdot \Bigl[ \int_1^{+\infty} exp(-2 
 \alpha^{-1} \xi r) \Bigl(1 + \frac{1}{2 \xi^2} \Bigr) \frac{\sqrt{\xi^2 - 1}}{\xi^2} d\xi \Bigr] 
 = \frac{Q}{r} + U(r) \label{phi}
\end{eqnarray}
where $\frac{Q}{r}$ is the Coulomb potential, $U(r)$ is the Uehling potential \cite{Uehl} and 
$\alpha = \frac{e^2}{\hbar c} = c^{-1}$ is the fine structure constant. Here and everywhere 
below in this study the notation $\hbar$ stands for the reduced Planck constant $\hbar = 
\frac{h}{2 \pi}$, or Dirac constant, $e$ designates the absolute value of the electric charge of 
the electron and $c$ is the speed of light in vacuum. In \cite{Our1} we have found the closed 
analytical formula for the Uehling potential (in atomic units)
\begin{eqnarray}
 U(r) = \frac{2 \alpha Q}{3 \pi r} \cdot \Bigl[ \Bigl(1 + \frac{r^2}{3 \alpha^2}\Bigr) 
 K_0\Bigl(\frac{2 r}{\alpha}\Bigr) - \frac{r}{6 \alpha} Ki_1\Bigl(\frac{2 r}{\alpha}\Bigr)
 - \Bigl(\frac{5}{6} + \frac{r^2}{3 \alpha^2}\Bigr) 
 Ki_2\Bigl(\frac{2 r}{\alpha}\Bigr) \Bigr] \label{UehA}
\end{eqnarray}
where $b = \frac{1}{\alpha}$ and the function $K_0(a)$ is the modified Bessel function of zero 
order (see Eqs.(8.432) and (8.447) from \cite{GR}), i.e. 
\begin{eqnarray}
 K_0(z) = \int_0^{\infty} exp(-z \cosh t) dt = \sum_{k=0}^{\infty} 
 (\psi(k+1) + \ln 2 - \ln z) \frac{z^{2 k}}{2^{2 k} (k!)^2} \; \; \; , 
 \label{macd}
\nonumber
\end{eqnarray}
where $\psi(k)$ is the Euler $psi$-function defined by Eq.(8.362) from \cite{GR}. The functions 
$Ki_1(z)$ and $Ki_2(z)$ in Eq.(\ref{UehA}) are the recursive integrals of the $K_0(z)$ function, 
i.e.
\begin{eqnarray}
 Ki_1(z) = \int_z^{\infty} Ki_0(z) dz \; \; \; , \; \; \; and \; \; \;
 Ki_n(z) = \int_z^{\infty} Ki_{n-1}(z) dz  \; \; \; , \label{repin}
\end{eqnarray}
where $n \ge 1$ and $Ki_0(z) \equiv K_0(z)$. After publication of \cite{Our1} we have found that 
the same problem was considered by Pauli and Rose in 1936 \cite{PauliRose}. They produced a 
similar expression for the Uehling potential (see the very last formula in their work). Recently, 
we have found a way to transform our formula, Eq.(\ref{UehA}), to the form which was derived by 
Pauli and Rose in \cite{PauliRose}. 

To concude this Section we note that the formula, Eq.(\ref{UehA}), can also be written in the form
\begin{eqnarray}
 U(z) &=& \frac{Q}{9 \pi z} \cdot \Bigl[ \Bigl( 12 + z^2 \Bigr) 
 K_0( z ) - z Ki_1( z ) - \Bigl( 10 + z^2 \Bigr) Ki_2( z ) \Bigr] \nonumber \\
 &=& \frac{Q}{9 \pi z} \cdot \Bigl[ \Bigl( 11 + z^2 - 1 \Bigr) K_0( z ) - z Ki_1( z ) - 
 \Bigl( 11 + z^2 + 1 \Bigr) Ki_2( z ) \Bigr] \label{UehAA} \\
 &=& \frac{Q}{9 \pi z} \cdot \Bigl[ \Bigl( q(z) - 1 \Bigr) K_0( z ) - z Ki_1( z ) - 
 \Bigl( q(z) + 1 \Bigr) Ki_2( z ) \Bigr] \nonumber
\end{eqnarray}
where $q(z) = z^2 + 11$ is a quadratic function of $z$. The tridiagonal form of the Uehling potential,
Eq.(\ref{UehAA}), is convenient to perform additional theoretical analysis based on methods developed 
in the theory of group representations.

\section{Wichmann-Kroll potential}

The formula, Eq.(\ref{UehA}), for the Uehling potential is corret mathematically, but it shows a wrong 
asymptotics at large interparticle distances $r$, e.g., at $r \ge 10 \alpha a_0$, where $a_0 = 
\frac{\hbar^2}{m_e e^2}$ is the Bohr radius. In general, any interaction potential which decays exponentially 
at large interparticle distances has a restricted physical meaning. On the other hand, it can be shown (see
below) that electric field generated by the higher order vacuum polarization correction to the Coulomb field 
has a power-type asymptotics at large $r$. Therefore, it is clear that at relatively large distances $r$ the 
overall contribution from this correction will overweight the correction from the Uehling potential. To 
explain the problem in detail let us consider the short- and long-range asymptotics of the Uehlling 
potential, Eq.(\ref{UehA}). First, note that the short range asymptotic of the $\phi(r)$ potential, 
Eq.(\ref{UehA}), takes the form 
\begin{equation}
 \phi(r)\mid_{r \rightarrow 0} \simeq \frac{Q}{r} \Bigl\{ 1 + \frac{\alpha}{3 \pi} \Bigl[ -\frac53 - 2 
 \gamma + 2 \ln \alpha - 2 \ln r \Bigr] \Bigr\} 
\end{equation}
where $\gamma \approx$ 0.577215664901532860606512$\ldots$ is the Euler constant (see, e.g., \cite{GR}) and $\ln \alpha 
\approx -4.92024365857$. The long-range asymptotics of the potential $\phi(r)$, Eq.(\ref{UehA}), is (in 
atomic units) 
\begin{equation}
 \phi(r)\mid_{r \rightarrow \infty} \simeq \frac{Q}{r} \Bigl\{ 1 + 
 \frac{\alpha^{\frac52}}{4 \sqrt{\pi} r^{\frac32}} \exp(-\frac{2}{\alpha} r) 
 \Bigr\} 
\end{equation}
This means that the long-range asymptotics of $\phi(r)$ decreases with $r$  vanishes at $r \rightarrow +\infty$ 
very rapidly $\sim r^{-\frac32} \exp(-\frac{2}{\alpha} r) \approx r^{-\frac32} \exp(-274 r)$. It has no direct 
physical sense for the corrections originated by a Coulomb interaction potential. Indeed, in this case we always
have a competing correction to the interaction (Coulomb) potential which is related to the lowest order vacuum 
polarization correction to the electromagnetic field $({\bf E}, {\bf H})$, or EM-field, for short. In actual atoms, 
the intensity of the magnetic field is much smaller than the intensity of the electric field of the nucleus, i.e. 
we can assume that $({\bf E}, {\bf H}) \approx {\bf E}$ and restrict our analysis below to the electric field only. 
It is shown below that this lowest order vacuum polarization correction to the electric field has power-type 
asymptotics at large interparticle distances. Briefly, this means that the electric filed created by the lowest 
order corrections to the vacuum polarization (which is produced by the electric field itself) decreases at 
$r \rightarrow +\infty$ much slower than the long-range tail of the Uehling potential, which decreases as
$\sim r^{-\frac32} \exp(-\frac{2}{\alpha} r) \approx r^{-\frac32} \exp(-274 r)$. Therefore, such a correction 
(known as the Wichmann-Kroll correction \cite{WK}) must always be taken into account when for realistic description 
of the vacuum polarization in actual atomic systems.  

Let us obtain the vacuum polarization correction of the lowest order to the (Coulomb) electric field ${\bf E}$ 
by considering one single electric charge $e$ (point) which is assumed to be at rest in the center of 
coordinates $x = 0, y = 0$ and $z = 0$. The corresponding Lagrangian $L$ which generates this correction
is written in the following form (in regular units) (see, e.g., \cite{AB})  
\begin{equation}
 L = \frac12 {\bf E}^2 + \frac{e^4 \hbar}{360 \pi^2 m_e^4 c^7} {\bf E}^4 = 
 \frac12 ({\nabla \phi})^2 + \frac{e^4 \hbar}{360 \pi^2 m_e^4 c^7} 
 ({\nabla \phi})^4 = L[\phi(r)] \label{LL}
\end{equation}
where ${\bf E} = -\nabla \phi(r)$. By varying the potential $\phi(r)$ in the following equation (the 
fundamental equation of least action \cite{WK}) 
\begin{equation}
  \delta \int L[\phi(r)] r^2 dr = 0 
\end{equation}
one finds the differential equation
\begin{equation}
 \frac{d \phi}{dr} + \frac{e^4 \hbar}{90 \pi^2 m_e^4 c^7} 
 \Bigl( \frac{d \phi}{dr} \Bigr)^3 = \frac{C}{r^2} \label{qubic}
\end{equation}
where $C = - \frac{Q e}{4 \pi}$. Assuming that $\phi(r) = \frac{Q e}{4 \pi r} + \psi(r)$, where $\psi(r)$ is 
a very small correction we reduce the last equation to the form
\begin{equation}
 \frac{d \psi}{dr} = \frac{e^4 \hbar}{90 \pi^2 m_e^4 c^7} 
 \Bigl( \frac{Q^3 e^3}{64 \pi^3 r^6} \Bigr)
\end{equation}
From this equation one finds 
\begin{equation}
 \psi(r) = -\frac{e^7 \hbar}{450 \pi^2 m_e^4 c^7} 
  \Bigl( \frac{Q^3}{64 \pi^3 r^5} \Bigr)
\end{equation}
Therefore, the total interaction potential $\phi(r)$ is
\begin{equation}
 \phi(r) = \frac{Q e}{4 \pi r} - \frac{e^7 \hbar}{450 \pi^2 m_e^4 c^7} 
 \Bigl( \frac{Q^3}{64 \pi^3 r^5} \Bigr) = 
 \frac{Q e}{4 \pi r} \Bigl[ 1 - \frac{2 \hbar}{225 \pi 
 m_e^4 c^7} \Bigl(\frac{Q^2 e^6}{64 \pi^3} \Bigr) \frac{1}{r^4} \Bigr] 
\end{equation}

In the relativistic units $\hbar = 1, c = 1$ and $\alpha = \frac{e^2}{4 \pi}$ the last formula takes the form
\begin{equation}
 \phi(r) = \frac{Q e}{4 \pi r} \Bigl[ 1 - \frac{2 Q^2 \alpha^3}{225 \pi 
 m_e^4 r^4} \Bigr] \label{eq11} 
\end{equation}
This expression is the sum of the usual Coulomb potential and a small correction which exactly coincides with the 
Wichmann-Kroll potential derived in \cite{WK}. In atomic units the expression, Eq.(\ref{eq11}), takes the form
\begin{equation}
 \phi(r) = \frac{Q}{r} \Bigl[ 1 - \frac{2 Q^2 \alpha^{7}}{225 \pi r^4} \Bigr] = \frac{Q}{r} - \frac{2 Q^3 
 \alpha^{7}}{225 \pi r^5}
\end{equation}
and Wichmann-Kroll potential $W_K(r)$ is
\begin{equation}
  W_K(r) = -\frac{2 Q^3 \alpha^{7}}{225 \pi r^5} \label{WK0}
\end{equation}
Note that: (1) this potential is always negative, and (2) it is singular at the origin. However, as follows from the 
Appendix such a singularity is pure formal, since Eq.(\ref{qubic}) is not correct at very small $r$. It is also 
clear that the corection produced by the Wichmann-Kroll potential at short distances will always be overweighted 
by the contribution from the Uehling potential. This means that the mentioned singularity of the Wichmann-Kroll potential, 
Eq.(\ref{WK0}), at $r \rightarrow 0$ can be removed, e.g., by using the replacement $r \rightarrow r + a$ which does 
not change other known properties of the Wichmann-Kroll potential. After some additional analysis of this situation 
and by using a number of trials we have found that the following `regularized' expression for the Wichmann-Kroll 
potential $W_K(r)$ (in atomic units)
\begin{equation}
  W_K(r) = -\frac{2 Q^3 \alpha^{7}}{225 \pi r (r^2 + \alpha^2)^2} \label{WK}
\end{equation}
has a number of advantages in actual applications to atomic and molecular systems. In particular, such a form of 
Wichmann-Kroll potential transforms this potential into a regular expression for atomic (or Coulomb) systems. It is 
also clear from Eq.(\ref{WK}) that at short interparticle distances the Wichmann-Kroll correction is not important 
and its contribution is significantly smaller than analogous contribution from the Uehling potential. On the other 
hand, at large distances, e.g. for $r \ge 10 \alpha a_0$ (and even for $r \ge \alpha a_0$) the relative contribution 
of the Wichmann-Kroll potential $W_K(r)$ to the sum $U(r) + W_K(r)$ is substantial and cannot be neglected in real 
atomic systems. It follows from the fact that the contribution from Uehling potential rapidly vanishes with $r$. 

Now, we can write the following general formula for the total interaction potential (in atomic units) between the two 
point electric charges ($Qe$ and $e$)
\begin{eqnarray}
 \Phi(r) = \frac{Q}{r} + U(r) + W_K(r) = \frac{Q}{r} + 
 \frac{2 Q \alpha}{3 \pi r} \cdot \Bigl[ \Bigl(1 + \frac{r^2}{3 \alpha^2}\Bigr) 
 K_0\Bigl(\frac{2 r}{\alpha}\Bigr) - \frac{r}{6 \alpha} 
 Ki_1\Bigl(\frac{2 r}{\alpha}\Bigr) \nonumber \\
 - \Bigl(\frac{5}{6} + \frac{r^2}{3 \alpha^2}\Bigr) 
 Ki_2\Bigl(\frac{2 r}{\alpha}\Bigr) \Bigr] 
 -\frac{2 Q^3 \alpha^{7}}{225 \pi r (r^2 + \alpha^2)^2} \label{total}
\end{eqnarray}
This interaction potential $\Phi(r)$ has correct asymptotic behaviour for both small and large interparticle distances 
$r$. In the case of interaction between two point electric charges $q_1 e$ and $q_2 e$ we need to replace in 
Eq.(\ref{total}) the factor $Q$ by the product $q_1 q_2$. The potential $\Phi(r_{12})$ takes the form (in atomic units)
\begin{eqnarray}
 \Phi(r_{12}) = \frac{q_1 q_2}{r_{12}} + 
 \frac{2 q_1 q_2 \alpha}{3 \pi r_{12}} \cdot \Bigl[ \Bigl(1 + \frac{r_{12}^2}{3 
 \alpha^2}\Bigr) K_0\Bigl(\frac{2 r_{12}}{\alpha}\Bigr) - \frac{r_{12}}{6 \alpha} 
 Ki_1\Bigl(\frac{2 r_{12}}{\alpha}\Bigr) \nonumber \\
 - \Bigl(\frac{5}{6} + \frac{r_{12}^2}{3 \alpha^2}\Bigr) 
 Ki_2\Bigl(\frac{2 r_{12}}{\alpha}\Bigr) \Bigr] 
 -\frac{2 (q_1 q_2)^3 \alpha^{7}}{225 \pi r_{12} (r^2_{12} + \alpha^2)^2} \label{totalt}
\end{eqnarray}
where $r_{12}$ is the distance between electrically charged particles 1 and 2. In regular units one needs to replace in 
Eqs.(\ref{total}) - (\ref{totalt}) the dimensionless fine structure constant $\alpha$ by the factor $\alpha = 
\frac{\hbar^2}{m_e e^2} = \alpha a_0$, where $a_0$ is the Bohr radius. 

\section{Fourier spatial resolution of the vacuum polarization potential}

In many problems known in nuclear, atomic and molecular physics and in Quantum Elecrodynamics one needs to use the 
closed analytical expression for the Fourier spectral resolution of the potential which describes the vacuum 
polarization corrections of the lowest order upon $\alpha$. As we have shown above such a potential is the sum of the
Uehling and Wichmann-Kroll potentials. Note that both the Uehling and Wichmann-Kroll potentials are the static and 
central potentials. Therefore, the Fourier spectral resolution is a superposition of plane waves of zero frequency, i.e.
in this case the spectral resolution is a spatial resolution. Moreover, it can be shown that all these plane waves are 
longitudinal, i.e. they are oriented along the spatial vector ${\bf k}$. First, consider the Fourier spatial resolution 
(or Fourier resolution, for short) of the Uehling potential which is written in the form
\begin{eqnarray}
 U({\bf r}) = \int^{+\infty}_{-\infty} \int^{+\infty}_{-\infty} 
 \int^{+\infty}_{-\infty} exp(-\imath {\bf k} \cdot {\bf r}) u({{\bf k}})
 \frac{d^3{\bf k}}{(2 \pi)^3} \label{eqf1} 
\end{eqnarray}
where ${\bf k} = (k_x, k_y, k_z)$ is the wave vector and $u_{{\bf k}}(r)$ is the unknown spectral function. Let us 
obtain the closed analytical formula for this spectral function and compare it with the spectral function of the pure 
Coulomb potential. 

From Eq.(\ref{eqf1}) one finds the following equation for the spectral function $u({{\bf k}}) = u(k)$
\begin{eqnarray}
 u(k) = \int \int \int U(2 b r) exp(\imath {\bf k} \cdot {\bf r}) d^3{\bf r} = 
 4 \pi \int^{+\infty}_0 j_0(k r) U(2 b r) r^2 dr \label{eqf2} 
\end{eqnarray}
where $b = \frac{1}{\alpha}$ and $k = \sqrt{k^2_x + k^2_y + k^2_z}$ is the radial component of the wave vector and 
$j_0(x)$ is the spherical Bessel function of zero order. Note also that the integration in the last formula is 
performed over the area occupied by the electric charges and electric field. In Eq.(\ref{eqf2}) we have used the 
fact that the Uehling potential is spherically symmetric and, therefore, we can integrate over all angular variables. 
By using Eq.(\ref{eqf1}) and expression for the $j_0(x) = \frac{sin x}{x}$ function one finds the following formula 
for the spectral function $u(k)$
\begin{eqnarray}
 u(k) = \frac{2 \alpha^3 Q}{3} \int_{1}^{+\infty} \frac{1}{t^2 + a^2} \Bigl( 1 + 
 \frac{1}{2 t^2} \Bigr) \frac{\sqrt{t^2 - 1}}{t^2} dt \label{eqf3} 
\end{eqnarray}
where $a = \frac{k}{2 b} = \frac{k \alpha}{2}$. The integral in the last equation can
be determined with the use of the substitution $u = \frac{t}{\sqrt{t^2 - 1}}$. It 
reduces the integral to the form
\begin{eqnarray}
 I = \int_{1}^{+\infty} \frac{1}{t^2 + a^2} \Bigl( 1 + \frac{1}{2 t^2} \Bigr) 
 \frac{\sqrt{t^2 - 1}}{t^2} dt = -\frac12 \int_1^{+\infty} \frac{1}{u^2 (1 + 
 a^2) - a^2} \cdot \Bigl( 3 - \frac{1}{u^2} \Bigr) \frac{du}{u^2} \label{eqf4} 
\end{eqnarray}
Then by using the substitution $y = \frac{1}{u}$ one transforms this integral to the
form
\begin{eqnarray}
 I = \frac{1}{2 a^2} \int_{0}^{1} \frac{(3 - y^2) y^2 dy}{c^2 - y^2} = \frac{1}{2 a^2}
 \Bigl[ -\frac53 + \frac{1}{a^2} - \frac12 \frac{\sqrt{a^2 + 1}}{a} \Bigl(2 - 
 \frac{1}{a^2}\Bigr) \ln \Bigl(\frac{\sqrt{1 + a^2} + a}{\sqrt{1 + a^2} - a}\Bigr) \Bigr]
 \label{eqf5} 
\end{eqnarray}
where $c^2 = \frac{a^2 + 1}{a^2}$. 

The final formula for the (spatial) spectral function $u(k)$ takes the form
\begin{eqnarray}
 u(k) =  \frac{\alpha^3 Q}{3 a^2} \Bigl[-\frac{5}{3} + \frac{1}{a^2} - 
 \frac12 \frac{\sqrt{a^2 + 1}}{a} \Bigl(2 - \frac{1}{a^2}\Bigr) \ln \Bigl(\frac{\sqrt{1 + 
 a^2} + a}{\sqrt{1 + a^2} - a}\Bigr) \Bigr] \label{eqf6} 
\end{eqnarray}
or, since $a = \frac{k \alpha}{2}$
\begin{eqnarray}
 u(k) =  \frac{8 \alpha Q}{3 k^2} \Bigl[-\frac{5}{6} + \frac{2}{\alpha^2 k^2} - 
 \frac{\sqrt{\alpha^2 k^2 + 4}}{\alpha k} \Bigl(1 - \frac{2}{\alpha^2 k^2}\Bigr) 
 \ln \Bigl(\frac{\sqrt{4 + \alpha^2 k^2} + k \alpha}{\sqrt{4 + \alpha^2 k^2} - k \alpha}\Bigr)
 \Bigr] \label{eqf65} 
\end{eqnarray}
As follows from the last formula the first term in the expression for $u(k)$ is $\sim k^{-2}$. 
This gives the asymptotics of the spectral function of the Uehling potential at large wave 
numbers $k$. The spectral function for the Coulomb potential between two interacting 
electrically charged particles is $u_C(k) = \frac{4 \pi}{k^2}$ (in atomic units) (see, e.g., 
\cite{LLE}). There are also a number of differences between the two spectral functions $u(k)$, 
Eq.(\ref{eqf65}), and $u_C(r)$. First, the $u(k)$ function contains various powers of the 
inverse wave number $k^{-1}$, while $u_C(k)$ is an exact quadratic function. Second, the $u(k)$ 
function, Eq.(\ref{eqf65}), contains inverse powers of the fine-structure constant $\alpha$, 
while the $u_C(k)$ function does not depend upon $\alpha$, if its expressed in atomic units. 
These two differences can also be found for the corresponding electric fields which are 
determined as the gradients of the corresponding potentials. The ${\bf k}$-component (Fourier 
component) of the total electric field (Coulomb + Uehling potentials) is ${\bf E}_{{\bf k}} = 
- \imath \frac{4 \pi}{k^2} (1 + \tilde{U}(k)) \cdot {\bf k}$, where the function $\tilde{U}(k)$ 
is
\begin{eqnarray}
 \tilde{U}(k) =  \frac{2 \alpha Q}{3 \pi} \Bigl[-\frac{5}{6} + \frac{2}{\alpha^2 k^2} - 
 \frac{\sqrt{\alpha^2 k^2 + 4}}{\alpha k} \Bigl(1 - \frac{2}{\alpha^2 k^2}\Bigr)
 \ln \Bigl(\frac{\sqrt{4 + \alpha^2 k^2} + k \alpha}{\sqrt{4 + \alpha^2 k^2} - k \alpha}\Bigr) 
 \Bigr] \label{eqf7} 
\end{eqnarray}
As follows from this formula each of the ${\bf E}_{{\bf k}}$ components is oriented along 
the wave vector ${\bf k}$, i.e. the total electric field contains only the longitudinal 
components (there are no non-zero transverse components) and these components do not depend 
upon the time $t$. Also, as one can see from the formula Eq.(\ref{eqf7}) in the lowest-order
approximation the contribution from the Uehling potential can be described as a small change
of the nuclear electric charge $Q \rightarrow Q \Bigl( 1 - \frac{5 \alpha}{9 \pi} \Bigr)$.  

As mentioned above the Uehling potential is a model potential which can be used to represent
the actual interparticle interactions only at relatively short distances. For larger distances
only the sum of the Uehling and Wichmann-Kroll potential must be considered. Let us obtain the
explicit formula for the Fourier spatial resolution of the Wichmann-Kroll potential. In general,
such a formula depends upon the regularization of the Wichmann-Kroll potential at $r \rightarrow 
0$. On the other hand, it is clear that at short and very short distances the correction which
corresponds to the  Wichmann-Kroll potential will be overweighted by the analogous correction 
generated by the Uehling potential. Briefly, this means that there is some freedom in the choice
of the regularization of the Wichmann-Kroll potential at $r \rightarrow 0$. Below, we chose the
Wichmann-Kroll potential in the form, Eq.(\ref{WK}). In this case the Fourier spatial resolution is:
\begin{eqnarray}
 w_K(k) &=& \int \int \int W_K(r) \exp(\imath {\bf k} \cdot {\bf r}) d^3{\bf r} = 
 4 \pi \int^{+\infty}_0 j_0(k r) W_K(r) r^2 dr \nonumber \\
 &=& - \frac{8 Q^3 \alpha^{7}}{225 k} \int^{+\infty}_0 \frac{r \sin(k r) dr}{(r^2 + \alpha^2)^2} 
  = - \frac{2 \pi Q^3 \alpha^{6}}{225} exp(-k \alpha) \label{eqf2wk} 
\end{eqnarray}
This allows one to obtain the following formula for the electric field ${\bf E}_{{\bf k}} = - 
\imath \frac{4 \pi}{k^2} (1 + \tilde{U}(k) + W_K(k)) \cdot {\bf k}$, where the function $\tilde{U}(k)$ 
is defined in Eq.(\ref{eqf7}), while the $W_K(k)$ function is
\begin{eqnarray}
   W_K(k) = - \frac{Q^3 \alpha^{6}}{450} k^2 exp(-k \alpha) \label{eqf3wk}
\end{eqnarray}

This result indicates clearly that the sum of the Uehling and Wichmann-Kroll potentials (i.e. the potential which 
is responsible for the lowest order correction on vacuum polarization) can be incorporated into the basic equations 
of Quantum Electrodynamics from the very beginning (see, e.g, \cite{Heitl} and \cite{AB}). Formally, it follows 
from the known fact that the longitudinal component of the electric field is not quantized \cite{Heitl}. Note that 
for the first time the formula for the Fourier spatial resolution of the Uehling potential (our Eq.(\ref{eqf65})) 
was also produced by Pauli and Rose in \cite{PauliRose} (see the formula given in footnote 4 in \cite{PauliRose}). 
It is interesting to note that the formulas derived in this Section can directly be used to determine the scattering 
amplitudes in the lowest-order plane-wave approximation (or Born approximation \cite{LLQ}). However, in many 
applications one needs to obtain more accurate approximations for the corresponding cross-sections. Another factor 
which drastically reduces the validity of the Born approximation is a substantial weakness of the both Uehling and 
Wichmann-Kroll potentials. This leads to the fact that the Coulomb potential cannot be ignored during such a 
scattering in any approximation. Briefly, this means that we need to apply more accurate procedures which produce 
reasonable answers. This problem is discussed in the next Section.

\section{Vacuum polarization correction to the Coulomb scattering}

The problem of Coulomb scattering is of paramount importance in a large number of applications known in modern physics. 
It is clear that the vacuum polarization of the Coulomb field will produce small changes in this well known process. In 
general, all possible changes in corresponding phase shifts are small ($\delta \sim \alpha$), but they can be important 
in some cases. For instance, the lowest order correction for vacuum polarization to the Coulomb scattering can be observed 
and measured in modern experiments. Here we derive the basic equations which describe such an `additional' scattering 
related with the vacuum polarization. Our approach is based on the variable phase method \cite{Babik}. In this method by 
using the known short-range interaction potential $U(r)$ we determine the corresponding variable phase $\delta_{\ell}(r)$ 
as a function of interparticle distance $r$. For two electrically charged particles interacting by a short-range, central 
potential $U(r)$ the phase function $\delta_{\ell}(r)$ is determined from the equation
\begin{eqnarray}
 \frac{d \delta_{\ell}(r)}{dr} = - \frac{2 U(r)}{k} \Bigl[ \cos\delta_{\ell}(r) F_{\ell}(k r; \eta) + 
 \sin\delta_{\ell}(r) G_{\ell}(k r; \eta) \Bigr]^2  \label{phase}
\end{eqnarray}
with the use of the initial condition $\delta_{\ell}(r = 0) = 0$. In Eq.(\ref{phase}) the notation $F_{\ell}(k r; \eta)$
stands for the the Coulomb functions regular at $r = 0$, while the notation $G_{\ell}(k r; \eta)$ designates the Coulomb 
functions which are singular at $r = 0$ (see, e.g., \cite{AS}). The parameter $\eta$ used in the formulas for the Coulomb 
functions is simply related with the electric charges $q_1 e$ and $q_2 e$ of the two interacting particles
\begin{eqnarray}
 \eta = \frac{e}{\hbar} \sqrt{\frac{m_1 m_2 q_1 q_2}{(m_1 + m_2) k}} \label{phase2}
\end{eqnarray}
where the parameter $k$ is wave number, i.e. $k = \sqrt{2 \frac{m_1 m_2 E}{(m_1 + m_2) \hbar^2}}$, and $E$ is the energy of the
colliding particles (in the center-of-mass). For atomic systems with an infinitely heavy nucleus and in atomic units ($\hbar 
= 1, m_e = 1$ and $e = 1$) one finds that $\eta = \sqrt{\frac{Q}{k}}$ and $k = \sqrt{2 E}$.

The Coulomb function of the first kind $F_{\ell}(k r; \eta)$ in Eq.(\ref{phase}) is simply related with the confluent 
hypergeometric function of the first kind ${}_1F_1(a;b;z)$ \cite{AS}
\begin{eqnarray}
 F_{\ell}(\rho; \eta) = C_{\ell}(\eta) \rho^{\ell+1} \exp(-\imath \rho) \cdot 
 {}_1F_1(\ell + 1 - i \eta; 2 \ell + 2; 2 \imath \rho)  \label{phase3}
\end{eqnarray}
where the factor $C_{\ell}(\rho)$ is
\begin{eqnarray}
   C_{\ell}(\eta) = \frac{2^{\ell} \exp(-\frac{\imath \eta}{2}) \mid \Gamma(\ell + 1 + \imath \eta) \mid}{\Gamma(2 \ell + 2)}
\end{eqnarray}
where $\Gamma(z)$ is the Euler $\Gamma-$function (see, e.g., \cite{GR}, \cite{AS}). Note that the Coulomb function 
$F_{\ell}(k r; \eta)$ is regular at $r = 0$.  The Coulomb function $G_{\ell}(k r; \eta)$ of the second type is 
\begin{eqnarray}
 G_{\ell}(\rho; \eta) =  \frac{2 \eta}{C^2_{0}(\eta)} \cdot {}_1F_1(\ell + 1 - i \eta; 2 \ell + 2; 2 \imath \rho) 
 \Bigl[ \ln(2 \rho) + \frac{q_{\ell(\eta)}}{p_{\ell}(\eta)} \Bigr] \label{phase4} \\
 + \frac{1}{(2 \ell + 1) C_{\ell}(\eta) \rho^{\ell}} \sum^{\infty}_{K=-\ell} a^{\ell}_{K}(\eta) \rho^{K+\ell} \nonumber 
\end{eqnarray}
where the values of all coefficients $q_{\ell(\eta)}, p_{\ell}(\eta)$ and $a^{\ell}_{K}(\eta)$ can be found in \cite{AS}. The Coulomb 
functions of the second kind are signular at $r = 0$. By solving the phase equation, Eq.(\ref{phase}), for the phase function   
$\delta_{\ell}(r)$ one can determine the long range asymptotics of the phase function $\delta_{\ell}(r = \infty) = \delta_{\ell}$ at
different wave numbers $k$ (or energies $E$). These values are needed to describe the actual interference between Coulomb and vacuum 
polarization scattering amplitudes.  

This result can be formulated in a different form. Let us consider the scattering of an electron at atomic nucleus with the electric 
charge $Q$ (or $Q e$) at small energies (or small wave numbers $k$). At such conditions the contribution from the term with $\ell = 0$ 
will substantially exceed analogous contribution from all other terms with $\ell \ge 1$. Therefore, below we can restrict ourselves to
the analysis of one term (with $\ell = 0$) only. The total scattering amplitude $f(\theta)$ is represented in the form 
\begin{eqnarray}
 f(\theta) = f_{C}(\theta) + \frac{1}{2 \imath k} \Bigl[ \exp(2 \imath \delta_{0}) - 1 \Bigr] \exp(2 \imath \delta^{C}_{0}) 
 \label{scat1}
\end{eqnarray}
where $f_{C}(\theta)$ and $\delta^{C}_{0}$ are the pure Coulomb scattering amplitude and Coulomb phase, respectively. The additional
phase $\delta_{0}$ corresponds to a small corrections for vacuum polarization. Briefly, the scattering amplitude contains two different
term (Coulomb term + vacuum polarization term). In the following formula for the total differential cross-section the both parts of 
scattering amplitude interfere with each other 
\begin{eqnarray}
 \frac{d\sigma}{d\theta} = \mid f(\theta) \mid^2 = \frac{Q^2}{2 v^2} \Bigl[ \frac{1}{\sin^{4}\frac{\theta}{2}} - 
 - \frac{4 k a_C}{\sin^{2}\frac{\theta}{2}} \sin\delta_{0} \cos\Bigl( \frac{2}{k a_C} \ln \sin\frac{\theta}{2} + \delta_{0} \Bigr)
 + 4 (k a_C)^2 \sin^2\delta_{0} \Bigr] \label{scat2}
\end{eqnarray}
where the parameter $a_C$ is related with the phase $\delta_{0}$ by the following equation
\begin{eqnarray}
 \cot\delta_{0} = -\frac{1}{\pi} \Bigl[ \exp\Bigl(\frac{2 \pi}{k a_C}\Bigr) - 1 \Bigr] \Bigl[ Re \psi\Bigl(-\frac{\imath}{k a_C}\Bigr)
  + \ln (k a_C) + \frac{\kappa a_C}{2} \Bigr] \label{scat3}
\end{eqnarray} 
where $\psi(z)$ is the logarithmic derivative of Euler's $\Gamma$-function, i.e. $\psi(z) = \frac{\Gamma^{\prime}(z)}{\Gamma(z)}$ , while 
the constant $\kappa$ is defined in \cite{LLQ}.

As follows from Eq.(\ref{scat2}) the interference between two parts of the scattering amplitude leads to the conclusion that differential
cross-section, Eq.(\ref{scat2}), is a linear function of the fine structure constant $\alpha$. Therefore, we can expect that effects of 
vacuum polarization in the electron-nuclear scattering will appear already in the lowest order approximation upon the fine structure
constant $\alpha$. Note in conclusion that the vacuum polarization effect can also be observed in high energy elastic scattering of fast 
electrons by few- and many-electron atoms. The corresponsing changes in the atomic form-factors are simply related with our formulas for
the Fourier spatial resolution of the vacuum polarization potential(s) derived above.   

\section{Cusp problem}

As it follows from the basics of atomic theory \cite{Fock1954} in any Coulomb few-body system the expectation value of the following 
operator 
\begin{eqnarray}
 \hat{\nu}_{ij} = \frac{1}{\langle \delta({\bf r}_{ij}) \rangle} \delta({\bf r}_{ij}) 
 \frac{\partial}{\partial r_{ij}} \label{eqc1}
\end{eqnarray}
computed with the bound state wave functions ($\Psi$) equals to the following value
\begin{eqnarray}
 \langle \Psi \mid \hat{\nu}_{ij} \mid \Psi \rangle = \Bigl( \frac{m_i m_j}{m_i + m_j} \Bigr) q_1 q_2
\end{eqnarray}
In particular, if $i = e$ (electron) and $j = N$ (nucleus), then the expectation value of the $\hat{\nu}_{ij}$ operator from Eq.(\ref{eqc1}) 
equals $-Q$ exactly (in atomic units). The coincidence of the expectation value of this operator with its expected value, i.e. with $-Q$, is 
often used in numerical computations to test the overall quality of the trial atomic wave functions. Also, in all atomic and polyelectron 
systems the expectation value of the electron-electron cusp always equlas $\frac12$. The coincidence between expected and actual 
electron-electron cusps is another test for the approximated wave function. Note also that the electron-nucleus and electron-electron cusps 
are of great interest for the general theory of bound states in Coulomb few-body systems.  

Now, consider the case when the sum of the Uehling $U(r)$ and Wichmann-Kroll $W_K(r)$ potentials is added to the Coulomb potential into the 
Schr\"{o}dinger equation. This will lead to the change of the variational wave function. The cusp condition between each pair of interacting, 
electrically charged  particles will change correspondingly. The fundamental problem is to find an analytical expression for the modified 
cusp, i.e. for the $\nu_{ij} = \langle \hat{\nu}_{ij} \rangle$ expectation value for an arbitrary $N$-electron atom. Formally, the answer can 
be written in the following form (in atomic units)
\begin{eqnarray}
 \nu_{eN} = - Q \Bigl\{ 1 - \frac{\alpha}{3 \pi} \bigl[ \frac53 + 2 \gamma - 
 2 \ln \alpha + 2 \ln r \Bigr] \Bigr\} + f \cdot \frac{2 Q^3 \alpha^3}{225 \pi} \label{eqc2}
\end{eqnarray}
where $\alpha$ is the fine structure constant, $f$ is some numerical factor and $\gamma$ is the 
Euler's constant. Note that the explicit expression for the $\nu_{eN}$ expectation value, Eq.(\ref{eqc2}), is $r-$dependent and its limit at 
$r \rightarrow 0$ does not exist. On the other hand, the definition of the cusp includes the two-particle delta-function, Eq.(\ref{eqc1}). For 
non-relativistic hydrogen atom this delta-function is a constant at all distances shorter than $a_{min} = \Lambda_e = \alpha a_0 = \alpha$ (in 
atomic units), where $\Lambda_e$ is the Compton wavelength. For atomic systems with $Q \ge 2$ the explicit expression for minimal distance 
becomes $Q-$dependent, e.g., $a_{min} = \frac{\Lambda_e}{Q}$. 

It is clear that the non-relativistic wave function cannot produce the actual electron density distribution at distances shorter than $a_{min} 
= \frac{\Lambda_e}{Q}$ (in atomic units). Instead it always gives the constant value. Briefly, this means that in Eq.(\ref{eqc2}) we have to 
assume that $a_{min} = C \alpha a_0 = C \alpha$ in Eq.(\ref{eqc2}), where $C$ is a numerical constant which is uniformly related 
to the electron-nuclear (and electron-electron) delta-function. Therefore, the cusp expectation value (or $\nu_{ij}$ value) is now finite and its 
numerical value is
\begin{eqnarray}
 \nu_{ij} =  - Q \Bigl\{ 1 - \frac{\alpha}{3 \pi} \bigl[ \frac53 + 2 \gamma + 2 \ln C \Bigr] \Bigr\} 
 + \frac{2 Q^3 \alpha^3}{225 \pi (1 + C^2)} \label{eqc4}
\end{eqnarray}
In other words, the Uehling potential produces a small correction to the pure Coulomb cusp which is $\sim \alpha$, while the correction which 
corresponds to the Wichmann-Kroll potential is even smaller $\sim \alpha^3$ (but it increases rapidly with the nuclear charge $Q$). It will be 
very interesting to perform highly accurate computations for some few-electron atoms with the mixed interaction potential between electrically 
particles (Coulomb + Uehling + Wichmann-Kroll). Such potentials must be directly introduced into the non-relativistic Schr\"{o}dinger equation. 
The expectation values of the actual electron-nuclear and electron-electron cusp values will be different from the values known for pure Coulomb 
systems. In modern highly accurate computations the cusp values are determined to the accuracy $\approx 1 \cdot 10^{-12}$ $a.u.$ The expected 
Uehling correction to the pure Coulomb cusp is $\approx 1 \cdot 10^{-3}$ (or 0.1 \%) and it can easily be detected in such computations. The 
Wichmann-Kroll potential also contribute to the electron-nucleus cusp, but such a contribution is significantly smaller $\approx 1 \cdot 10^{-7}$ 
$a.u.$ (for light atoms with relatively small nuclear charge $Q \le 5$).

\section{Consecutive approximations upon the fine structure constant}

As mentioned above the short-range Uehling potential can be considered as a small correction to the Coulomb potential. Therefore, the contributions 
of the Uehling potential can be taking into account with the use of the perturbation theory. The small parameter in this theory is the fine structure 
constant $\alpha$ (or the factor 2 $\alpha$). This fact is well known from earlier studies, but for atomic (or Coulomb) systems we can perform such an 
perturbation analysis based on the dynamical O(2,1)-algebra of the three radial operators written below. This algebara is responsible for separation of 
the discrete and continuous spectra (see, e.g., \cite{Arnd}, \cite{Bar}). This means that by using the commutation relations between three generators 
of this O(2,1)-algebra we can separate the discrete and continuous spectra from each other. 

In this study we apply the known commutation relations between generators of this radial O(2,1)-algebra to develop a closed procedure to  evaluate 
all consecutive contributions of the vacuum polarization potential upon the fine structure constant $\alpha$. Below in this Section, we restrict 
ourselves to the analysis of the Coulomb two-body systems. The cases of the Coulomb three- and four-body systems can be considered analogously. The 
Coulomb two-body system (or one-electron atom) includes an infinitely heavy nucleus with the electric charge $Qe$ and an electron of the mass $m_e$ 
and electric charge $-e$. The corresponding Schr\"{o}dinger equation (in atomic units) takes the 
following form 
\begin{eqnarray}
 \hat{H} \Psi = \Bigl[ -\frac12 \Bigl( \frac{\partial^2}{\partial r^2} + \frac{2}{r} 
 \frac{\partial}{\partial r} \Bigr) + \frac{\hat{L}^2}{r^2} - \frac{Q}{r} \Bigr] \Psi 
 = E \Psi \label{Eq1} 
\end{eqnarray}
where $\hat{H}$ is the Hamiltonian, $E$ is the energy of the system, $\hat{L^2}$ is the operator of the angular momentum of the system (atom) and $\Psi$ is 
the wave function. The operator $\hat{L^2}$ depends only upon two angular variables ($\Theta$ and $\phi$) and it commutes with any operator which depends 
upon the radial variable $r$. The explicit form of the $\hat{L^2}$ operator is
\begin{eqnarray}
 \hat{L^2} = - \Bigl[ \frac{1}{\sin\Theta} \frac{\partial}{\partial \Theta} \Bigl(
 sin\Theta \frac{\partial}{\partial \Theta} \Bigr) + \frac{1}{\sin^2\Theta}
 \frac{\partial^2}{\partial \phi^2} \Bigr]
\end{eqnarray}
The eigenfunctions of the $\hat{L^2}$ operator are the spherical harmonics (see, e.g., \cite{AB}) $Y_{\ell m}(\Theta,\phi)$ and $\hat{L^2} Y_{\ell 
m}(\Theta,\phi) = \ell (\ell + 1) Y_{\ell m}(\Theta,\phi)$.
  
In \cite{Bar} (see also \cite{Fro821} and \cite{Fro85}) it was shown that three following operators
\begin{eqnarray}
  S = \frac12 r \Bigl( p^2_r + \frac{\hat{L}^2}{r^2} + 1 \Bigr) \; \; \; , \; \; \;
  T = r p_r \; \; \; \; \; and \; \; \; \; \;
  U = \frac12 r \Bigl( p^2_r + \frac{\hat{L}^2}{r^2} - 1 \Bigr) \label{Eq3}
\end{eqnarray}
form the non-compact O(2,1)-algebra with the commutation relations
\begin{eqnarray}
  [ S, T ] = -\imath U \; \; \; , \; \; \; [ T, U ] = \imath S \; \; \; , \; \; \;
  [ U, S ] = -\imath T \label{Eq4}
\end{eqnarray}
The Casimir operator of this algebra $C_2 = S^2 - U^2 - T^2$ equals to the $\hat{L}^2$ operator of the angular momentum. The exact coincidence of the Casimir 
operator $C_2$ of the radial algebra $O(2,1)$ with the Casimir operator $C_2 = L^2$ of the compact algebra of three-dimensional rotations $O(3)$ is related 
with the complementarity of the corresponding representations in the space of more general $Sp(6 N,R)$-algebra representations. Such a complementarity of 
the `radial' and `angular' representations defined by Moshinsky and Quesne in \cite{Mos}.  

Based on this fact we can transform the operator $r (\hat{H} - E) = [(S + U) - E (S - U) + Q]$ from the Schr\"{o}dinger equation by applying the unitary 
transformation $exp(\imath \beta T)$, where $\beta$ is the real parameter and $T$ is the generator from Eq.(\ref{Eq3}). Indeed, by using the well known 
Hausdorff formula
\begin{eqnarray}
  exp(-\imath \beta T) (S \pm U) exp(\imath \beta T) = exp(\pm \beta T) (S \pm U) \label{Eq5}
\end{eqnarray}
and choosing $\beta = ln\sqrt{-2 E}$ we reduce the original Schr\"{o}dinger equation to the form
\begin{eqnarray}
  [ \sqrt{- 2 E} S - Q ] \phi(r,\Theta,\phi) = 0 \label{Eq6}
\end{eqnarray}
The operator $S$ has the discrete spectrum only \cite{Bar}, \cite{Kog}, since $S \mid n, \ell, m \rangle = n \mid n, \ell, m \rangle$, where $n = 1, 2, 3, 
\ldots$ is an integer positive number. From Eq.(\ref{Eq6}) one finds $E_n = -\frac{Q^2}{2 n^2}$ (the energy bound state spectrum in atomic units) and $\mid n, 
\ell, m \rangle = A \phi_{n}(r) Y_{\ell m}(\Theta, \phi)$, where $A$ is an arbitrary numerical constant, $\phi_{n}(r)$ are the radial functions and 
$Y_{\ell m}(\Theta, \phi)$ are the spherical harmonics. All these facts and equations are very well known for the Coulomb two-body systems (atoms) (see, e.g., 
\cite{Bar}). It appears that we can modify this procedure for pure Coulomb systems to include both the Uehling potential and Wichmann-Kroll potnetials. 

Here we want to discuss very briefly an approach which is used for such a generaliztion. First, we note that from Eq.(\ref{Eq3}) it follows that $r = S + U$. 
Second, the modified Uehling potential $r U(r)$ is written in the form:
\begin{eqnarray}
 r U(r) = \frac{2 \alpha Q}{3 \pi} \Bigl[ \Bigl(1 + \frac{b^2 r^2}{3}\Bigr) K_0(2 b r) - \frac{b r}{6} Ki_1(2 b r) - \Bigl(\frac{b^2
 r^2}{3} + \frac{5}{6}\Bigr) Ki_2(2 b r) \Bigr] \; \; \; , \label{UehX}
\end{eqnarray}
This means that the Uehling potential is an analytical function of the $2 b r = \frac{2}{\alpha} (S + U)$ variable. Such a variable contain the small 
dimensionless parameter $b$ or $\frac{2}{\alpha}$, where $\alpha$ is the fine structure constant. Since $\alpha \approx \frac{1}{137}$ is samll, then $b = 
\frac{2}{\alpha}$ is a large dimmensionless parameter. In Eq.(\ref{UehX}) the parameter $\frac{2}{\alpha}$ plays the role of the cut-off parameter, since all 
modified Bessel functions $K_0(2 b r), Ki_1(2 b r)$ and $Ki_2(2 b r)$ decrease exponentially at large $r$. Briefly, this means that we can restrict the power 
series of the $r U(2 b r)$ potential to a very few first terms and expectation values of all these terms can be determined with the use of the commutation 
relations for the O(2,1)-algebra, Eq.(\ref{Eq4}). In particular, we have found that in the lowest order approximation the effect of vacuum polarization can be 
represented as a small change of the electric charge (pure Coulomb screening).  

\section{Conclusion}

We have derived the general formula for the interaction potential between two point electric charges which includes the lowest order correction for vacuum 
polarization. Our formulas, Eqs.(\ref{total}) - (\ref{totalt}), for such a potential agree with the lowest order QED-approximation and provide the correct asymptotic 
behaviour at arbitrary interparticle distances. The formulas, Eqs.(\ref{total}) - (\ref{totalt}), can directly be used in highly accurate computations of the bound 
states in few-electron atoms and ions. Vacuum polarization in few-electron atoms and ions leads to small changes ($\sim \alpha$) in the cusp values. Such changes are 
small but they can be determined with the use of highly accurate (atomic) wave functions obtained currently for many Coulomb few-body systems. We also discuss the 
role of the vacuum polarization correction for the electron-nucleus scattering. The corresponding effect $(\sim \alpha)$ can also be measured to high accuracy in 
modren experiments.    
 
\begin{center}
    {\bf Appendix}
\end{center}

The equation, Eq.(\ref{qubic}), from the main text is a cubic equation in respect to the radial derivative of the unknown potential $\phi(r)$. It 
can be re-written in the form  
\begin{equation}
 \frac{e^4 \hbar}{90 \pi^2 m_e^4 c^7} y^3 + y + \frac{Q e}{4 \pi r^2} = 0 
 \label{qub1}
\end{equation}
where $y = \frac{d \phi}{dr}$. As follows from the general theory of cubic functions this equation has only one real root (the discriminant of this 
equation is negative). In fact, this equation is reduced to the form of a monic trinomial (see, e.g., \cite{AS})
\begin{equation}
 y^3 + p y + \frac{q}{r^2} = 0 \label{qub2}
\end{equation}
where $p = \frac{90 \pi^2 m_e^4 c^7}{e^4 \hbar}$ and $q = \frac{45 \pi Q m_e^4 c^7}{2 e^3 \hbar}$ (in usual units). 

The Cardano method gives the only real root of Eq.(\ref{qub2})
\begin{equation}
 y = \sqrt[3]{-\frac{q}{2 r^2} + \sqrt{\frac{q^2}{4 r^4} + \frac{p^3}{27}}} + \sqrt[3]{-\frac{q}{2 r^2} - 
 \sqrt{\frac{q^2}{4 r^4} + \frac{p^3}{27}}} \label{root}
\end{equation}
Therefore, in the main text we obtain the following (unique) differential equation
\begin{equation}
 \frac{d \phi}{dr} = \sqrt[3]{-\frac{q}{2 r^2} + \sqrt{\frac{q^2}{4 r^4} + \frac{p^3}{27}}} + \sqrt[3]{-\frac{q}{2 r^2} 
 - \sqrt{\frac{q^2}{4 r^4} + \frac{p^3}{27}}} \label{root2}
\end{equation}
At $r \rightarrow \infty$ this equation has the solution $\phi(r) = 2 (\frac{p}{3})^{\frac12} \cdot r + c$, where $c$ is some numerical constant. At $r 
\rightarrow 0$ the analogous solution takes the form $\phi(r) = -3 q^{\frac13} \cdot r^{\frac13} + c$, where $c$ is some numerical constant. Both these 
asymptotics do not have any physical sense. Therefore, our substitution $r \rightarrow r + \alpha$ used in the main text to regularize the Wichmann-Kroll 
potential at small $r$ does not contradict any actual property of the $\phi(r)$ potential.

\end{document}